\documentclass[aps,prb,reprint]{revtex4-1}
\usepackage{amsmath}
\usepackage{graphicx}
\usepackage{color}
\usepackage{cases}
\usepackage{hyperref}

\newcommand{\fref}[1]{Fig.~\ref{#1}}

\newcommand{\pref}[1]{(\ref{#1})}

\newcommand{\eref}[1]{Eq.~\pref{#1}}

\begin{document}
\title{Spatial adiabatic passage processes in sonic crystals with linear defects}
\author{R.~Menchon-Enrich, J.~Mompart and V.~Ahufinger}
\affiliation{Departament de F\'{\i}sica, Universitat Aut\`{o}noma de Barcelona, E-08193 Bellaterra, Spain\\}
\date{\today}

\begin{abstract}
We investigate spatial adiabatic passage processes for sound waves propagation in sonic crystals, consisting of steel cylinders embedded in a water host medium, that present two linear defects. This work constitutes an extension of the well-known quantum optical rapid adiabatic passage technique to the field of sound propagation. Several spatial adiabatic passage devices are proposed, by appropriately designing the geometry of the two linear defects along the propagation direction, to work as a coherent multifrequency adiabatic splitter, a phase difference analyzer and a coherent multifrequency adiabatic coupler. These devices are robust in front of fluctuations of the geometric parameter values.
\end{abstract}

\pacs{PACS}

\maketitle
\section{Introduction}
A unique feature of the adiabatic passage processes, which  consist in the adiabatic following of an eigenvector of a system, is their working robustness
against fluctuations of the parameter values. Adiabatic passage processes have been successfully studied in several areas of physics, such as Quantum Optics, Ultracold Atoms, and light propagation in coupled waveguide systems. In Quantum Optics, two techniques have been extensively studied, the so-called stimulated Raman adiabatic passage \cite{bergmann_coherent_1998} (STIRAP) and the rapid adiabatic passage \cite{leslie_allen_optical_1987, shore_theory_1990} (RAP). In both cases, an eigenstate of an atomic system interacting with a specific sequence of laser pulses is modified in time and adiabatically followed, achieving a complete and robust transfer of  population between two internal atomic levels of the system. Adiabatic passage processes have been extended to the external degrees of freedom of Ultracold Atoms,\cite{eckert_three-level_2004} being proposed for the transport of single atoms between the most distant traps of a system of three tunneling-coupled potential wells, the so-called spatial adiabatic passage. 
Subsequently, spatial adiabatic passage processes have also been discussed for the transport of electrons,\cite{greentree_coherent_2004} Bose--Einstein condensates (BECs)\cite{graefe_mean-field_2006, rab_spatial_2008, cole_spatial_2008, ottaviani_adiabatic_2010} and holes,\cite{benseny_atomtronics_2010} for state \cite{loiko_filtering_2011} and velocity\cite{yurii2013} filtering of neutral atoms and also to transfer\cite{mcendoo_2010} and generate angular momentum states.\cite{rycka2013} 

Spatial adiabatic passage for light propagation in a system of three total internal reflection (TIR) coupled optical waveguides has been experimentally reported.\cite{longhi_coherent_2007, menchon-enrich_adiabatic_2012} In this case, a supermode of a triple-waveguide system is adiabatically modified and followed along the propagation direction, achieving a complete light transfer between the outermost waveguides of the system. Applications of the light spatial adiabatic passage, such as a polychromatic beam splitter \cite{dreisow_polychromatic_2009} and a spectral filter,\cite{menchon-enrich_light_2013} have been successfully experimentally demonstrated. Light guiding in waveguides has been broadly studied, being TIR waveguides the most technologically developed.\cite{chen_foundations_2007} However, the introduction of photonic crystals (PhCs) \cite{eli_yablonovitch_inhibited_1987, krauss_two-dimensional_1996,joannopoulos_photonic_2008} provided a new way of light guiding by means of linear defects, consisting in rows with unitary cells with a modified geometry,\cite{meade_photonic_1991, meade_novel_1994, mekis_high_1996, joannopoulos_photonic_1997} which could be integrated in much smaller sizes than the traditional TIR waveguides. Linear defects allow for light guiding due to the creation of propagation bands within the band gap frequencies. 

In analogy with the PhCs, phononic crystals (PCs) were introduced \cite{sigalas_band_1993, kushwaha_acoustic_1993, kushwaha_theory_1994, montero_de_espinosa_ultrasonic_1998, vasseur_experimental_2001} in the field of sound waves propagation, leading to numerous new physical phenomena for sound waves such as negative refraction and focusing,\cite{zhang_negative_2004, yang_focusing_2004} nondiffractive propagation,\cite{perez-arjona_theoretical_2007, espinosa_subdiffractive_2007, soliveres_self_2009} or angular band gaps,\cite{romero-garcia_angular_2013} to cite a few. As well, sound guiding in linear defects in PCs has been studied,\cite{kafesaki_frequency_2000, khelif_transmittivity_2002, khelif_transmission_2003, miyashita_full_2002, khelif_trapping_2003, khelif_guiding_2004, vasseur_absolute_2008} including systems of coupled linear defects.\cite{sun_analyses_2005}

In this work, spatial adiabatic passage processes for sound waves propagation in linear defects are addressed for the first time to the best of our knowledge. We will extend the well-known quantum optical rapid adiabatic passage technique to sound propagation to propose novel sonic devices working as coherent multifrequency adiabatic splitters, phase difference analyzers and coherent multifrequency adiabatic couplers. In particular, we will investigate the sound propagation in systems of two linear defects in sonic crystals (SCs).\cite{martinez-sala_sound_1995, sanchez-perez_sound_1998} SCs are a particular case of PCs that consist of solid scatterers embedded in a fluid host medium. SCs permit considering the propagation of only longitudinal waves, which constitutes an important simplification. Furthermore, SCs are experimentally relevant since, for example, they allow for measurements inside the crystal.\cite{romero-garcia_angular_2013} 

This paper is organized as follows. In Section \ref{sc_sectionSC} we will present the considered physical system consisting of two coupled linear defects in a SC, and we will calculate the allowed bands into the band gap and their corresponding supermodes. We will also distinguish between two different frequency ranges inside the band gap, one where two supermodes coexist, and another range with only one supermode available. In Section \ref{sc_twosmode}, for the frequency range in which both supermodes exist, we will study spatial adiabatic passage processes through the coupled-mode equations, and we will design a coherent multifrequency adiabatic splitter, in Subsection \ref{sc_CMAS}, and a phase difference analyzer, in Subsection \ref{sc_PDA}. In Section \ref{sc_onesmode}, we will focus on the frequency range with only one supermode, for which we will design spatial adiabatic passage processes that will make the system behave as a coherent multifrequency adiabatic splitter and as a coupler. Finally, in Section \ref{sc_CONC}, we will present the conclusions.

\section{Physical system}\label{sc_sectionSC}
We consider a two-dimensional square lattice SC made of steel cylinders (filling medium) immersed in water (host medium) containing linear defects. In the following, we will assume that the steel cylinders are a rigid material, disregarding transversal waves in the cylinders. This approximation, although it can slightly shift the bands in frequency,\cite{li_bandgap_2011} it is commonly used to describe the qualitative behavior of those processes which do not critically depend on a particular set of parameter values, and/or to show the existence of new physical phenomena.\cite{espinosa_subdiffractive_2007, soliveres_self_2009} In this context, the propagation of sonic waves in SCs consisting of two different materials can be described by the following linear equations:\cite{perez-arjona_theoretical_2007}
\begin{subequations}\begin{align}
\rho\frac{\partial \mathbf{v}}{\partial t}&=-\nabla p,\\
\frac{\partial p}{\partial t}&=-B\mathbf{\nabla\cdot v},
\end{align}\label{sc_linearized}\end{subequations}
where $B(\mathbf{r})$ is the bulk modulus, $\rho(\mathbf{r})$ is the density, $p(\mathbf{r},t)$ is the pressure field and $\mathbf{v}(\mathbf{r},t)$ is the velocity vector field. Merging Eqs.~(\ref{sc_linearized}a) and (\ref{sc_linearized}b) it is possible to obtain the wave equation for the pressure field describing the sound propagation in an inhomogeneous medium:
\begin{equation}
\frac{1}{B}\frac{\partial^2 p}{\partial t^2}- \nabla \cdot\left( \frac{1}{\rho}\nabla p\right)=0.
\label{sc_waveeq}
\end{equation}
Considering sound beams with harmonic temporal dependence and Fourier expanding $B(\mathbf{r})$ and $\rho(\mathbf{r})$ since they are periodic functions with 
the periodicity of the lattice, equation (\ref{sc_waveeq}) can be solved by means of the plane wave expansion (PWE) method, which gives an eigenvalue problem equation:\cite{kushwaha_acoustic_1993, kushwaha_theory_1994, perez-arjona_theoretical_2007}
\begin{equation}
\sum_{\mathbf{G}'}[\omega^2b_{\mathbf{G}-\mathbf{G}'}^{-1}-\rho_{\mathbf{G}-\mathbf{G}'}^{-1}(\mathbf{k}+\mathbf{G})\cdot(\mathbf{k}+\mathbf{G}')]p_{\mathbf{k},\mathbf{G}'}=0,
\label{sc_eigenvalue}
\end{equation}
where $\omega$ is the angular frequency of the plane waves divided by the sound velocity in the host medium, $c_h$, $\mathbf{k}$ is a two-dimensional Bloch vector belonging to the irreducible Brillouin zone, $\mathbf{G}$ and $\mathbf{G}'$ are vectors of the reciprocal lattice, $p_{\mathbf{k},\mathbf{G}'}$ is the coefficient for the pressure field expanded following the Bloch-Floquet theorem 

\begin{equation}
p(\mathbf{r})=e^{i\mathbf{k}\cdot\mathbf{r}}\sum_{\mathbf{G}'}p_{\mathbf{k},\mathbf{G}'}e^{i\mathbf{G}'\cdot\mathbf{r}},
\label{sc_bloch}
\end{equation}
and $b_{\mathbf{G}-\mathbf{G}'}^{-1}$ and $\rho_{\mathbf{G}-\mathbf{G}'}^{-1}$ are the Fourier coefficients of the inverted relative values of the bulk modulus, $\bar{B}(\mathbf{r})^{-1}=B_h/B(\mathbf{r})$, and the density, $\bar{\rho}(\mathbf{r})^{-1}=\rho_h/\rho(\mathbf{r})$, respectively. Here, $B_h$ and $\rho_h$ are the bulk modulus and the density of the host medium, respectively.

In the case of a SC consisting of only two different materials, the values of the coefficients $b_{\mathbf{\bar{G}}}^{-1}$ and $\rho_{\mathbf{\bar{G}}}^{-1}$, with $\mathbf{\bar{G}}=\mathbf{G}-\mathbf{G}'$, can be found by integrating over the filled area (corresponding to steel) inside the two-dimensional unit cell. For $\rho_{\mathbf{\bar{G}}}^{-1}$ one obtains \cite{kushwaha_acoustic_1993, kushwaha_theory_1994, perez-arjona_theoretical_2007}
\begin{numcases}{\rho_{\mathbf{\bar{G}}}^{-1}=}
\frac{1}{A}\iint\frac{1}{\bar{\rho}(\mathbf{r})}\mathrm{d}\mathbf{r}=\frac{\rho_h}{\rho_c}f+(1-f),&\hspace{-0.4cm}for $\mathbf{\bar{G}}=0$\hspace{0.5cm} \label{sc_coeff1}\\
\frac{1}{A}\iint\frac{e^{(i\mathbf{\bar{G}}\cdot\mathbf{r})}}{\bar{\rho}(\mathbf{r})}\mathrm{d}\mathbf{r},&\hspace{-0.4cm}for $\mathbf{\bar{G}}\neq0$\hspace{0.5cm}     \label{sc_coeff2}
\end{numcases}
where $A$ is the area of the unit cell, $f$ is the filling factor, and $\rho_c$ is the density of the filling material. In the case of a square lattice and a SC formed by cylinders of radius $r_0$, \eref{sc_coeff2} takes the following form:
\begin{equation}
\rho_{\mathbf{\bar{G}}}^{-1}=\left(\frac{\rho_h}{\rho_c}-1\right)2f\frac{J_1(|\mathbf{\bar{G}}|r_0)}{|\mathbf{\bar{G}}|r_0}, \,\rm{\ for\ } \mathbf{\bar{G}}\neq0,
\end{equation}
where $J_1$ is the Bessel function of the first kind. The expressions for $b_{\mathbf{\bar{G}}}^{-1}$ can be obtained analogously.\cite{perez-arjona_theoretical_2007}

The eigenvalues of \eref{sc_eigenvalue} can be numerically obtained and correspond to the frequencies of the allowed propagation bands in the SC. Once the frequencies are known, by using \eref{sc_bloch} one can calculate the supermodes of the structure, i.e., the pressure field $p(\mathbf{r})$.

\begin{figure}[!ht]
\centerline{
\includegraphics[width=0.9\linewidth]{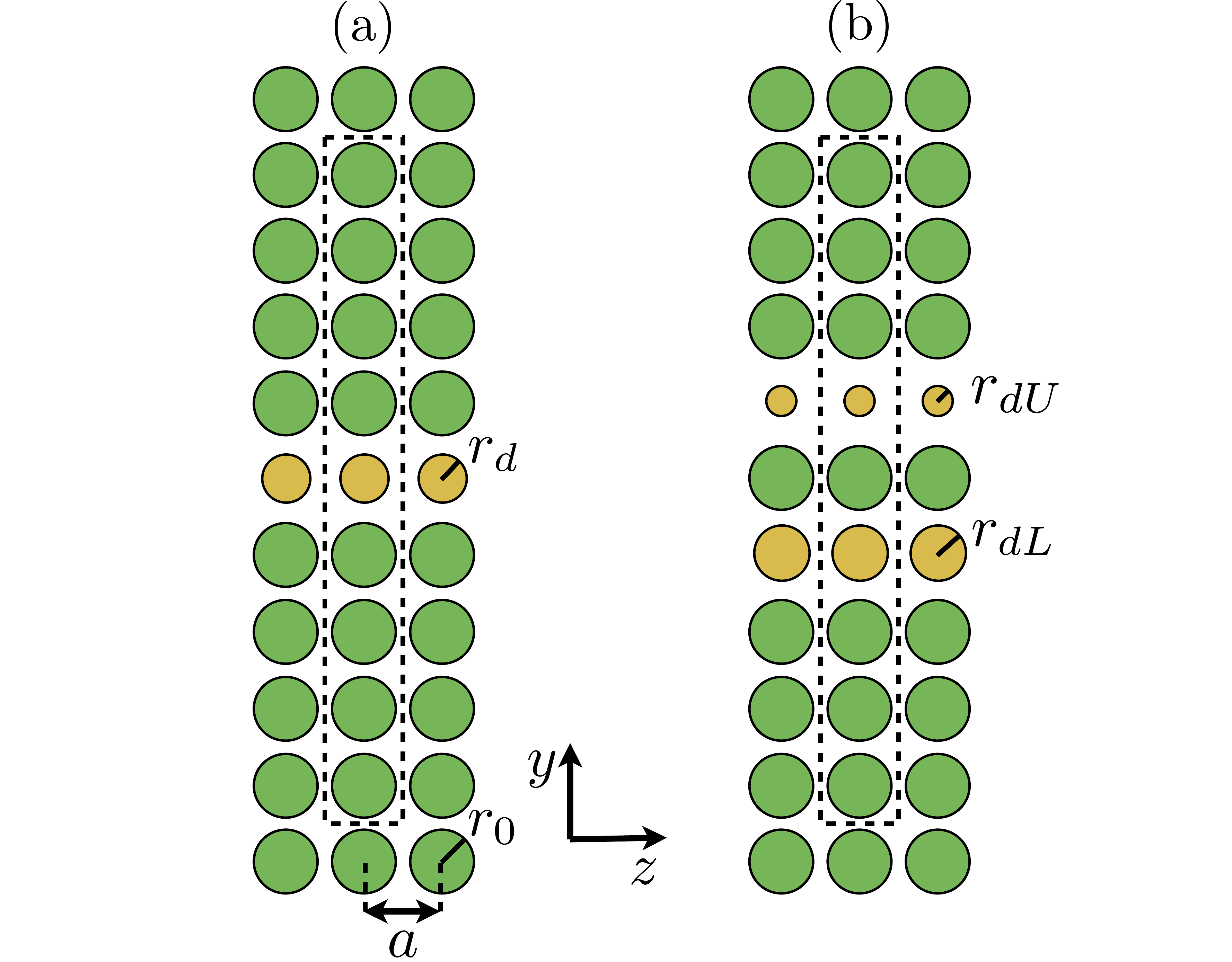}}
\caption{(Color online) Schematic representation of the square lattice SC with (a) one linear defect and (b) two linear defects separated by a single row. The supercells are boxed with a dashed line. The SC considered in this work has a lattice constant of $a=5\,\rm{mm}$ and consists of steel cylinders of radius $r_0=2.25\,\rm{mm}$ ($\rho_c=7.8\times10^3\,\rm{kg\ m^{-3}}$ and $B_c=160\times10^9\,\rm{N\ m^{-2}}$) immersed in water ($\rho_h=10^3\,\rm{kg\ m^{-3}}$, $B_h=2.2\times10^9\,\rm{N\ m^{-2}}$).}
\label{sc_supercells}
\end{figure}

\begin{figure*}[!ht]
\centerline{
\includegraphics[width=0.935\linewidth]{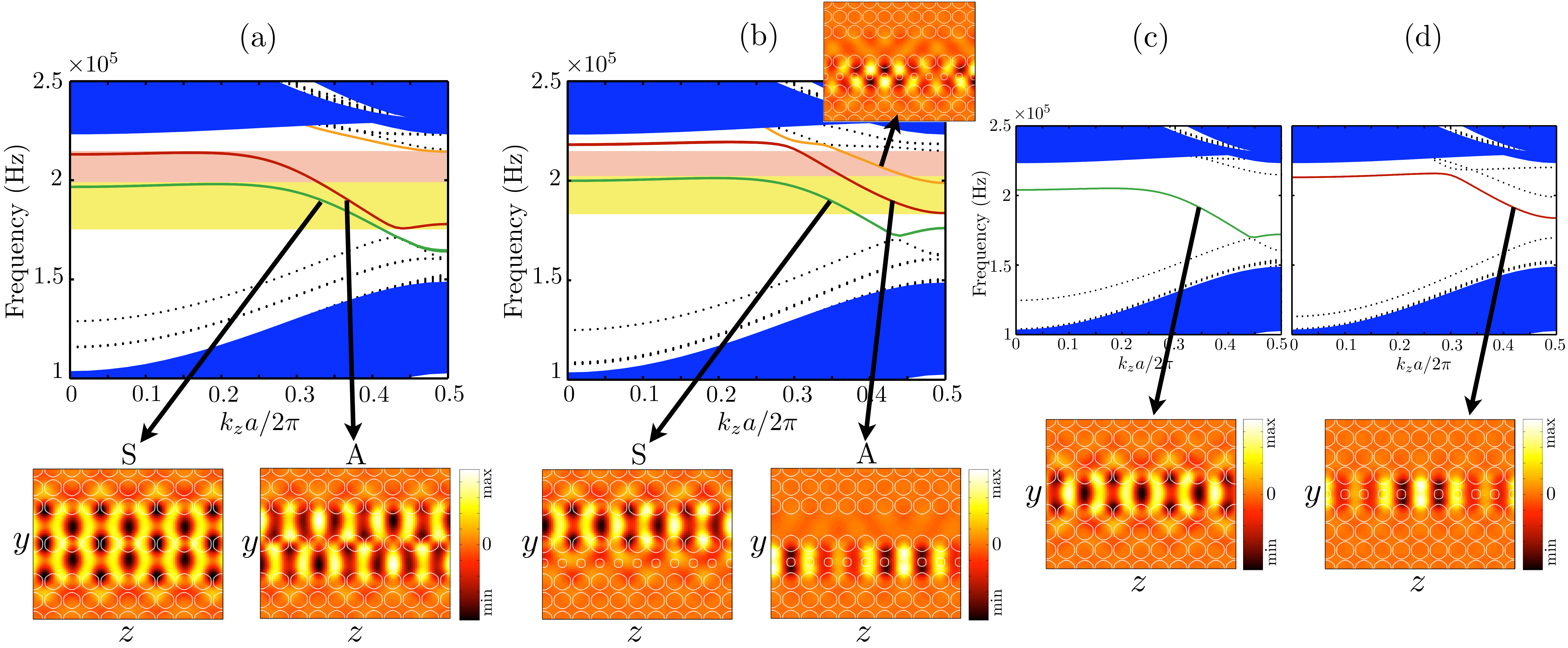}}
\caption{(Color online) Projected band diagrams for two linear defects of radii (a) $r_{dU}=r_{dL}=0$ and (b) $r_{dU}=0$ and $r_{dL}=1.2\,\rm{mm}$, and for a single linear defect of radius (c) $r_{d}=0$ and (d) $r_{d}=1.2\,\rm{mm}$, in a SC with the same parameter values as in \fref{sc_supercells}. In (a) and (b) figures, we show the S band, as a dark gray (green in the online version) solid line, and the A band, as a black (red in the online version) solid line, that will be used for the control of sound propagation. The light gray (orange in the online version) solid line band corresponds to a higher order supermode. Due to its antisymmetric transverse profile within one waveguide, this higher order supermode will not be excited by the considered input source with a Gaussian transverse pressure profile and it will not be taken into account in the rest of this work. The supermodes for the S and A bands are shown for a frequency $1.9\times 10^5\,\rm{Hz}$ by plotting the pressure field. The higher order supermode is also shown in (b) for a frequency $2.05\times 10^5\,\rm{Hz}$. In (a) and (b) the frequency regions where both the S and A supermodes coexist are marked with a light gray shadow (light yellow in the online version), and where only the A supermode exists is marked with a gray shadow (light red in the online version). (c) and (d) figures allow for the comparison of both, the projected band diagrams and the supermodes, corresponding to a single linear defect with the two-linear-defect case. The supermodes in (c) and (d) are represented for a frequency $1.9\times 10^5\,\rm{Hz}$. In all the projected band diagrams the dotted lines correspond to additional bands that will not be investigated in this work.}
\label{sc_fig2}
\end{figure*}

A way to create linear defects in SCs consists in adding rows of cylinders with different radii to the ones forming the square lattice. If we introduce one linear defect composed of cylinders of radius $r_d$ (see \fref{sc_supercells}(a)), bands corresponding to the modes of the individual waveguide are obtained into a frequency band gap in the projected band diagram of the SC, allowing for sound guiding inside the SC. Similarly, if we introduce two parallel linear defects with radii $r_{dU}$ (for the upper one) and $r_{dL}$ (for the lower one) separated, for example, by one row of cylinders of radius $r_0$ (see \fref{sc_supercells}(b)), bands corresponding to different  supermodes of the system appear into the band gap of the projected band diagram. For the calculation of the projected band diagrams of the structures with linear defects it is necessary to change the unit cell of the SC to a supercell, which contains several of the previous unit cells in the $y$ direction, as shown in \fref{sc_supercells}. The PWE method can also be used in this case.\cite{vasseur_absolute_2008} However, $b_{\mathbf{\bar{G}}}^{-1}$ and $\rho_{\mathbf{\bar{G}}}^{-1}$ have to be recalculated for the corresponding supercell 
with \eref{sc_coeff2}. 
For the case of one linear defect in a supercell containing nine cylinders (\fref{sc_supercells}(a)), \eref{sc_coeff2} becomes:
\begin{align}
\rho_{\mathbf{\bar{G}}}^{-1}=&\left(\frac{\rho_h}{\rho_c}-1\right)\frac{2\pi}{9a^2}\left[r_d\frac{J_1(|\mathbf{\bar{G}}|r_d)}{|\mathbf{\bar{G}}|}+\right. \notag
\\
&\left.2r_0\frac{J_1(|\mathbf{\bar{G}}|r_0)}{|\mathbf{\bar{G}}|}\sum_{j=1}^{4}{\cos(j\bar{G}_ya)}\right],\notag
\\
& \,\rm{\ for\ } \mathbf{\bar{G}}\neq0.
\label{sc_coeff2_a}
\end{align}
On the other hand, for the case of two defects separated by a single row in a supercell containing nine cylinders (\fref{sc_supercells}(b)), \eref{sc_coeff2} reads:
\begin{align}
\rho_{\mathbf{\bar{G}}}^{-1}=&\left(\frac{\rho_h}{\rho_c}-1\right)\frac{2\pi}{9a^2}\notag
\\
&\left[r_{dU}\frac{J_1(|\mathbf{\bar{G}}|r_{dU})}{|\mathbf{\bar{G}}|}e^{i\bar{G}_ya}+r_{dL}\frac{J_1(|\mathbf{\bar{G}}|r_{dL})}{|\mathbf{\bar{G}}|}e^{-i\bar{G}_ya}+\right.\notag
\\
&\left. 2r_0\frac{J_1(|\mathbf{\bar{G}}|r_0)}{|\mathbf{\bar{G}}|}\left(1/2+\sum_{j=2}^{4}{\cos(j\bar{G}_ya)}\right)\right],\notag
\\
 &\,\rm{\ for\ } \mathbf{\bar{G}}\neq0.
\label{sc_coeff2_b}
\end{align}
The corresponding expressions for $b_{\mathbf{\bar{G}}}^{-1}$ are analogous to \eref{sc_coeff2_a} and \eref{sc_coeff2_b}.

\fref{sc_fig2} shows the projected band diagrams obtained using the PWE method for (a) a SC with two equal defects corresponding to the absence of cylinders ($r_{dU}=r_{dL}=0$), (b) a SC with two defects of very different radius, $r_{dU}=0$ (one empty row) and $r_{dL}=1.2\,\rm{mm}$, (c) a SC with a single empty defect (of radius $r_d=0$) and (d) a SC with a single defect of $r_{d}=1.2\,\rm{mm}$. In \fref{sc_fig2} we also represent the supermodes for the most relevant allowed bands, which have been calculated using \eref{sc_bloch}. In this work we will focus in two of the obtained bands for the case of two linear defects (Figs.~\ref{sc_fig2}(a) and (b)), one represented as a dark gray solid line (green in the online version) which we will call S (as it corresponds to a symmetric supermode when both linear defects are of equal radius), and one represented as a black solid line (red in the online version) which we will call A (as it corresponds to an antisymmetric supermode when both defects are equal). It is worth to note that in \fref{sc_fig2}(b) there is another band (light gray solid line or orange color in the online version) which significantly overlaps in frequencies with the A band. This band corresponds to a higher order supermode of the system with an antisymmetric transverse profile within one waveguide (see \fref{sc_fig2}(b)). Thus, this supermode will not be excited by the  sound source that will be used, which has a symmetric transverse profile centered at the linear defect (a Gaussian transverse pressure profile). Figs.~\ref{sc_fig2}(c) and (d) allow for the comparison of both, the projected band diagrams and the supermodes, for a single and a two linear defect case. We have plotted one of the bands as a dark gray solid line (green in the online version) in \fref{sc_fig2}(c) and one as a black solid line (red in the online version) in \fref{sc_fig2}(d), similarly as in Figs.~\ref{sc_fig2}(a) and (b).

Let's continue now with the two linear defect case. In \fref{sc_fig2}(a) we can see that when the two defects are equal, both supermodes (S and A) spread equally into the two defects. However, when the defects have significant different radii (\fref{sc_fig2}(b)), the S supermode is localized in the linear defect with smallest radius ($r_{dU}=0$ in the figure) whereas the A supermode stays localized in the linear defect with larger radius ($r_{dL}=1.2\,\rm{mm}$ in the figure). In general, the transverse amplitude profiles of the supermodes can be modified by changing the difference between the radii sizes: the more different the radii of two defects are, the more localized the supermodes are in one of the defects. Thus, if one of the supermodes of the system is excited, either the S or the A, by smoothly changing the radii of the linear defects along the propagation direction it is possible to adiabatically follow the initially excited supermode of the system, which, due to the radii modification, will smoothly change its transversal amplitude profile along the propagation direction.\cite{johnson_adiabatic_2002} This adiabatic following, which corresponds to the so-called spatial adiabatic passage process,\cite{eckert_three-level_2004, greentree_coherent_2004, graefe_mean-field_2006, rab_spatial_2008, cole_spatial_2008, ottaviani_adiabatic_2010, benseny_atomtronics_2010, longhi_coherent_2007, menchon-enrich_adiabatic_2012} allows for the control of the sound propagation along linear defects in SCs. If the radius modification is not smoothly performed, sound waves will not be able to adiabatically follow the change of the transverse profile of the initially excited supermode along the propagation direction, exciting other supermodes of the system and we would lose the control of sound propagation. It is important to remark that spatial adiabatic passage processes have the advantage that they are robust in front of variations of the parameter values. In our case, fluctuations in the size of the defects are not critical to perform a spatial adiabatic passage process, only a smooth enough variation of the radii size is needed to adiabatically follow the considered supermode.

\section{Two-supermode case}\label{sc_twosmode}
The propagation of sound along two defects for frequencies in which both S and A supermodes exist (light gray shadow in Figs.~\ref{sc_fig2}(a) and (b), or yellow in the online version) can be approximately described by the coupled-mode equations, as it can be done for TIR waveguides \cite{chen_foundations_2007} or photonic crystal linear defects.\cite{chien_tight-binding_2007} For the most general case of two coupled waveguides, coupled-mode equations can be written as:
\begin{equation}
i\frac{d}{dz}\left(\begin{array}{c}C_{U}(z)\\C_{L}(z)\end{array}\right)= \frac{1}{2}\left( \begin{array}{cc}
2k_U (z) & -\Omega_{LU} (z)\\
-\Omega_{UL} (z) &2k_L (z)
\end{array} \right)\left(\begin{array}{c}C_{U} (z)\\C_{L} (z)\end{array}\right),
\label{sc_coupled}
\end{equation}
where $k_{U}$ ($k_{L}$) is the propagation constant of the individual upper (lower) waveguide, $\Omega_{UL}$ ($\Omega_{LU}$) is the coupling coefficient from the upper to the lower (from the lower to the upper) waveguide, and $C_{U}$ ($C_{L}$) represents the amplitude function of the field in the upper (lower) waveguide. For the case of sound, the total pressure field in a system of two coupled sonic waveguides can be expressed as
\begin{equation}
p(y,z)=\sum_{i=U,L}C_i(z)p_i(y),
\end{equation}
where $p_U$ ($p_L$) is the mode of only the upper (lower) waveguide. The velocity field can be described similarly.

It is possible to diagonalize \eref{sc_coupled} in order to obtain the expressions for the pressure field corresponding to the S supermode, $p_+(z)$, and the A supermode, $p_-(z)$, as a function of the parameters of the individual linear defects:
\begin{equation}
p_+(z)=\left (\begin{array}{c}\cos\theta/\sqrt{\frac{\Omega_{UL}}{\Omega_{LU}}\sin^2\theta+\cos^2\theta}\\\sqrt{\frac{\Omega_{UL}}{\Omega_{LU}}}\sin\theta/\sqrt{\frac{\Omega_{UL}}{\Omega_{LU}}\sin^2\theta+\cos^2\theta}\end{array}\right),
\label{sc_sim}
\end{equation} 
\begin{equation}
p_-(z)=\left (\begin{array}{c}\sin\theta/\sqrt{\sin^2\theta+\frac{\Omega_{UL}}{\Omega_{LU}}\cos^2\theta}\\-\sqrt{\frac{\Omega_{UL}}{\Omega_{LU}}}\cos\theta/\sqrt{\sin^2\theta+\frac{\Omega_{UL}}{\Omega_{LU}}\cos^2\theta}\end{array}\right),
\label{sc_antisim}
\end{equation}
where the mixing angle $\theta$ is defined by 
\begin{equation}
\tan{2\theta}=\frac{\sqrt{\Omega_{UL}\Omega_{LU}}}{\Delta k},
\label{sc_mixing}
\end{equation}
with $\Delta k=k_L-k_U$. The propagation constants of the $p_+(z)$ and $p_-(z)$ supermodes are 
\begin{equation}
k_{\pm}=\frac{k_U+k_L\mp\sqrt{\Omega_{UL}\Omega_{LU}+{\Delta k}^2}}{2}.
\label{sc_kmodes}
\end{equation}

In order to modify, along the propagation direction, the $p_+$ and $p_-$ supermodes obtained from the coupled-mode equations, it is necessary to change the couplings and thus, the mixing angle $\theta$ (see Eqs. (\ref{sc_sim}) and (\ref{sc_antisim})), which depends on the ratio $\sqrt{\Omega_{UL}\Omega_{LU}}/{\Delta k}$, as shown in \eref{sc_mixing}. This ratio, and therefore the supermode transverse profiles, can be controlled by modifying the radii of the defects. 
Thus, the coupled-mode equations (\ref{sc_coupled}) allow us to easily confirm that, as we have discussed in Section \ref{sc_sectionSC}, by smoothly changing the radii of the linear defects it is possible to modify and follow the S and A supermodes of the system and control the sound propagation. Calculating the projected band diagrams for a single defect with \eref{sc_eigenvalue}, \eref{sc_coeff1} and \eref{sc_coeff2_a} it is straightforward to check that the value of $|\Delta k|=|k_L-k_U|$ increases by making the defect radii progressively different between them (see Figs.~\ref{sc_fig2}(c) and (d)). In the case in which the defects are equal, $r_{dU}=r_{dL}$, then $\Delta k=0$. The values of the couplings $\sqrt{\Omega_{UL}\Omega_{LU}}$ can be found by using \eref{sc_kmodes}, since $k_\pm$ can be obtained from the band diagrams with two defects in the SC. Thus, it can be checked that when the radii of the defects are equal, $|\Delta k|\ll\sqrt{\Omega_{UL}\Omega_{LU}}$ and $\theta=\pi/4$. Whereas when $|r_{dL} - r_{dU}|\gg0$, the condition $|\Delta k|\gg\sqrt{\Omega_{UL}\Omega_{LU}}$ is fulfilled and $\theta$ is equal to either $0$ or $\pi/2$, depending on the sign of $\Delta k$. In this way, by changing the difference between the radii of the linear defects, it is possible to modify the mixing angle $\theta$ from either $0$ or $\pi/2$, corresponding to the S and A supermodes localized in only one of the defects, to $\pi/4$, where the S and A supermodes are equally spread between the two linear defects. This is in agreement with the results shown in Figs.~\ref{sc_fig2}(a) and (b).

Additionally, the coupled-mode equations provide a useful analogy between propagation of waves along two coupled waveguide systems and the time evolution of the population in two-level atomic systems interacting with a laser beam. In fact, the protocol to modify and follow the S and A supermodes presented here resembles the well-known quantum-optical rapid adiabatic passage (RAP) technique,\cite{leslie_allen_optical_1987, shore_theory_1990} which allows for a coherent control of atomic population in a system of two internal atomic states interacting with a chirped laser pulse.

\subsection{Coherent multifrequency adiabatic splitter}\label{sc_CMAS}
\begin{figure}[htb]
\centerline{
\includegraphics[width=1\linewidth]{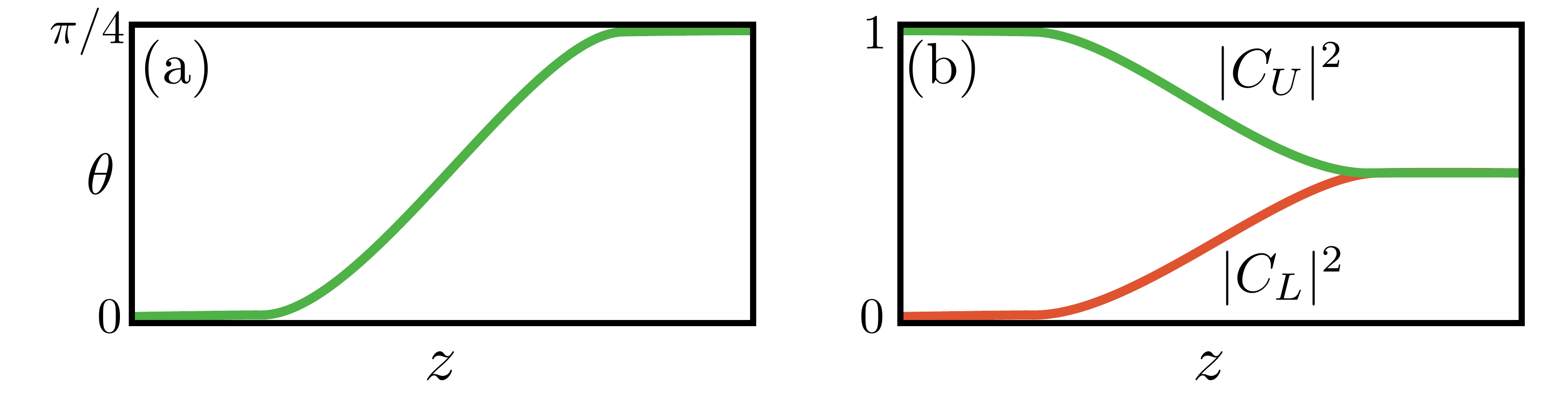}}
\caption{(Color online) Schematic representation of the evolution of (a) the mixing angle $\theta$ and (b) the field intensity in the upper and lower defects.}
\label{sc_migrap}
\end{figure}

In this section we will apply the previous ideas coming from the coupled-mode theory in order to obtain a robust $50\%$ coherent superposition of sound waves at the output of a SC with two coupled linear defects.
To this aim, we need to modify the mixing angle $\theta$ from either $0$ or $\pi/2$ to $\pi/4$. For example: if we consider that at the input of the system only the upper linear defect is excited ($C_U=1$ and $C_L=0$) and the radii are different enough so $\Delta k$ is large compared to $\sqrt{\Omega_{UL}\Omega_{LU}}$, then $\theta=0$ and only the S supermode $p_+(z_{\rm{initial}})$ is excited. Thus, if along the $z$ propagation direction the defect radii become progressively more similar (reaching equal values at the output of the system), $\Delta k$ decreases making $\sqrt{\Omega_{UL}\Omega_{LU}}$ large compared to $\Delta k$, and $\theta$ evolves adiabatically from $0$ up to $\pi/4$. At the output of the system, where the linear defects are equal and $\theta=\pi/4$, the followed S supermode $p_+(z_{final})$ in \eref{sc_antisim} corresponds to have $50\%$ of the field in each of the two linear defects. \fref{sc_migrap} schematically represents the evolution of the mixing angle $\theta$ and the 
amplitude in each linear defect provided by the coupled-mode equations. As summarized in table~\ref{sc_taula2}, there are four possible ways to achieve the $50\%$ power superposition at the output of the two coupled linear defects.

\begin{table}[htdp]
\begin{center}
\begin{tabular}{|c|c|c|c|}
\hline
\textbf{Sign of} &\textbf{Injection} & \textbf{Mixing angle}& \textbf{Followed}\\
\textbf{$\Delta k=k_L-k_U$} & & \textbf{evolution}& \textbf{supermode}\\
\hline
$+$ & $C_U=1$, $C_D=0$ & $0\Rightarrow \pi/4$&$p_+(z)$\\
\hline
$+$ & $C_U=0$, $C_D=1$ & $0\Rightarrow \pi/4$&$p_-(z)$\\
\hline
$-$ & $C_U=1$, $C_D=0$ & $\pi/2\Rightarrow \pi/4$&$p_-(z)$\\
\hline
$-$ & $C_U=0$, $C_D=1$ & $\pi/2\Rightarrow \pi/4$&$p_+(z)$\\
\hline
\end{tabular}
\end{center}
\caption{Four possible situations for which, at the output of the device, the field injected in one of the linear defects ends up in a $50\%$ superposition between the two parallel coupled linear defects.}
\label{sc_taula2}
\end{table}

\begin{figure*}
\centerline{
\includegraphics[width=\linewidth]{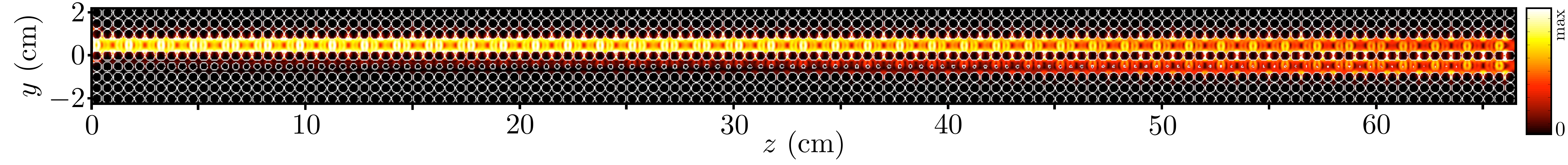}}
\caption{(Color online) Numerical simulation of the total intensity field for the propagation of a harmonic sound wave along the SC with two linear defects acting as a coherent beam splitter structure for a frequency of $1.9\times 10^5\,\rm{Hz}$. The parameter values of the SC are the same as in \fref{sc_supercells}. The upper linear defect consists of a row empty of cylinders whereas the lower linear defect smoothly changes its radius from $r_{dL}=2.25\,\rm{mm}$ to $r_{dL}=0$. The considered input source is a harmonic wave with a Gaussian transverse profile centered at the upper defect. Its full-width at half maximum for the pressure field is of $5\,\rm{mm}$ in order to approximately match the linear defect supermode profile.}
\label{sc_figsplit}
\end{figure*}

\begin{figure}[!ht]
\centerline{
\includegraphics[width=0.7\linewidth]{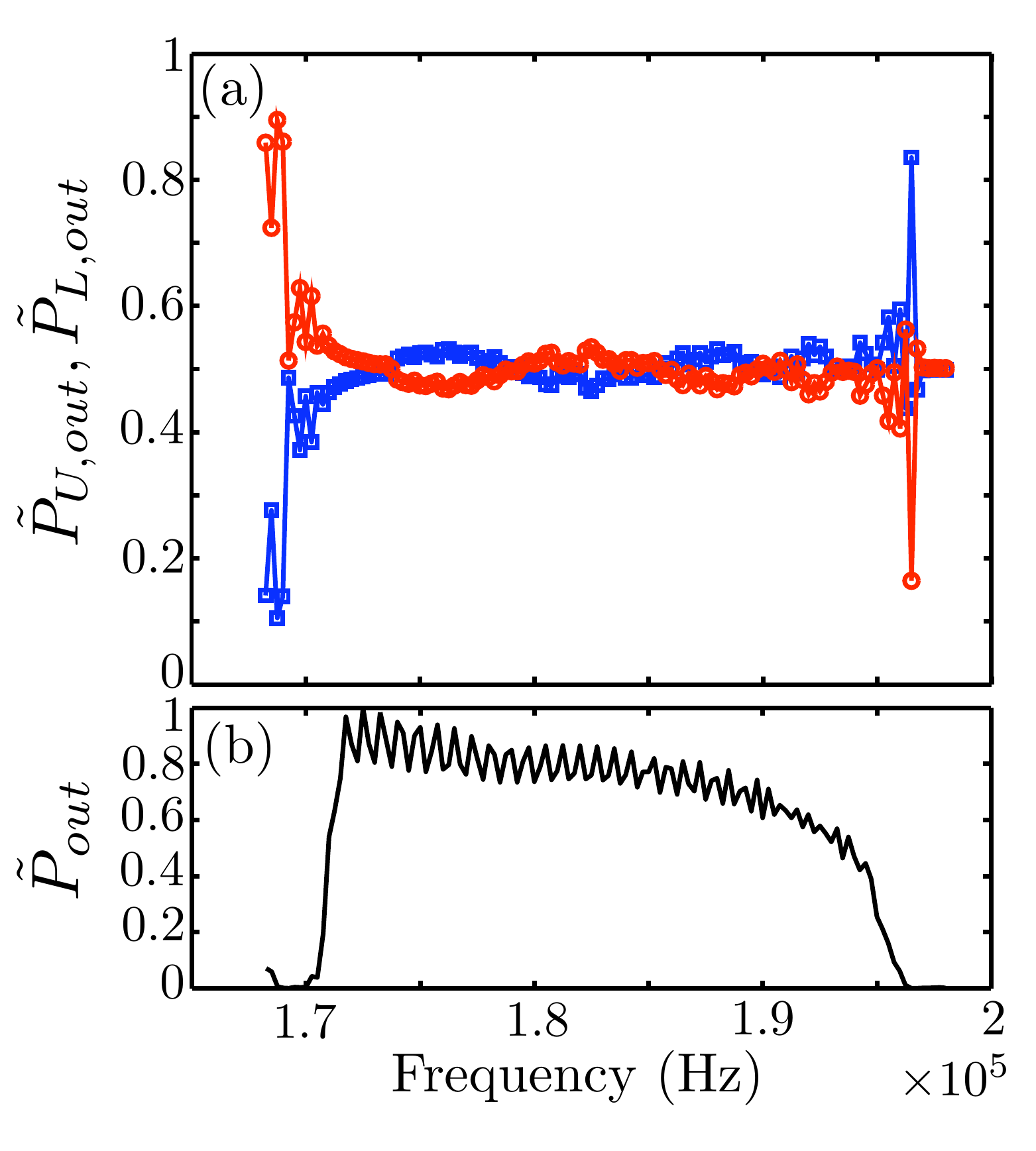}}
\caption{(Color online) (a) Normalized power at the upper $\tilde{P}_{U,out}$ (red dots) and lower $\tilde{P}_{L,out}$ (blue squares) outputs, and (b) normalized total power reaching the outputs, $\tilde{P}_{out}$, as a function of the frequency.}
\label{sc_figsplitrange}
\end{figure}

We have designed a structure following the example above described (corresponding to the first row of table~\ref{sc_taula2}) and performed numerical simulations by integrating \eref{sc_linearized} with the Finite Element method in order to confirm the predictions of the coupled-mode equations for the $50\%$ splitting. \fref{sc_figsplit} shows the results of the numerical simulation of sound waves of frequency $1.9\times 10^5\,\rm{Hz}$ propagating through the designed structure. We can see how the initial input beam injected in the upper defect is equally split into the two linear defects at the output. The input linear defect consists of an empty row and the second linear defect consists of cylinders that change their radius from $r_{dL}=2.25\,\rm{mm}$ (the size of the cylinders in the SC) to $r_{dL}=0$. The two linear defects are separated by one row of the SC.

Considering $P_{U,out}$ and $P_{L,out}$ as the power integrated over the width of the upper and the lower linear defects ($5\,\rm{mm}$), respectively, \fref{sc_figsplitrange}(a) plots the normalized power at the two outputs, $\tilde{P}_{U,out}=P_{U,out}/P_{out}$ and $\tilde{P}_{L,out}=P_{L,out}/P_{out}$ as a function of the frequency, where $P_{out}=P_{U,out}+P_{L,out}$. On the other hand, \fref{sc_figsplitrange}(b) shows the normalized total power reaching the outputs relative to the maximum power at the two outputs, i.e., $\tilde{P}_{out}=P_{out}/\max(P_{out})$. We can observe that the adiabatic splitting works for a significantly broad range of frequencies from $1.71\times 10^5\,\rm{Hz}$ to $1.95\times 10^5\,\rm{Hz}$, in agreement with the available frequencies for the S supermode in the band diagrams for the different positions along the propagation, see Figs.~\ref{sc_fig2}(a) and (b). Thus, this structure constitutes a coherent multifrequency adiabatic splitter. The splitting process works for different frequencies as long as it is possible to follow adiabatically the S supermode. If the supermode is not adiabatically followed, for example because of a too sudden change of the radius of the defect, the A supermode (or even some other higher supermodes) would be excited, limiting the efficiency of the splitting. As long as the adiabaticity condition is fulfilled, the length of the device is not a critical parameter in order to maintain an equal splitting. For longer devices as compared to the one here proposed, the adiabaticity of the process will also be successfully fulfilled and the splitting will be efficient for all the range of frequencies of the S supermode. However, for shorter devices, some oscillations in the splitting efficiency might start appearing as a result of the impossibility to adiabatically follow the S supermode.

\subsection{Phase difference analyzer}\label{sc_PDA}
\begin{figure*}[!ht]
\centerline{
\includegraphics[width=0.66\linewidth]{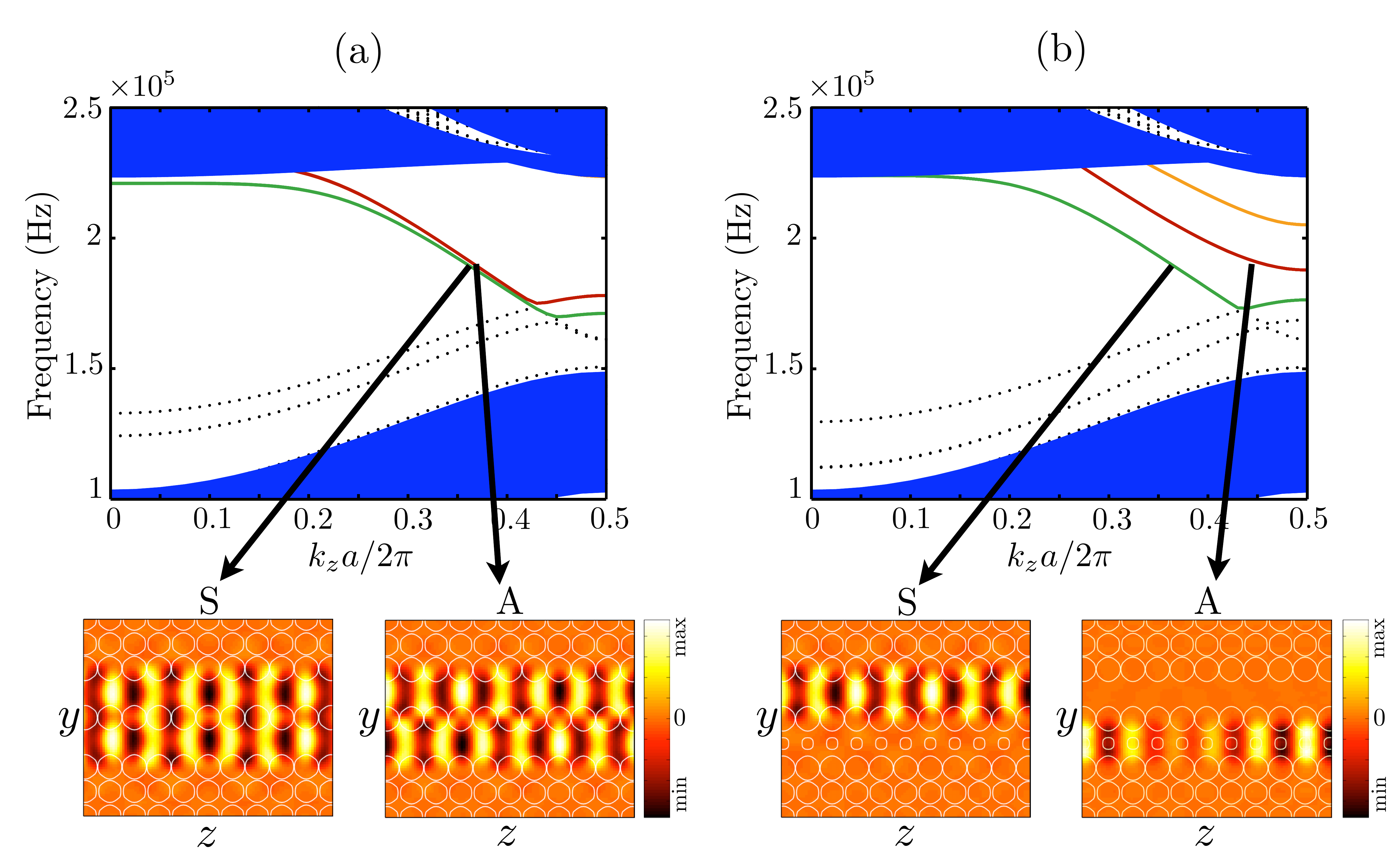}}
\caption{(Color online) Projected band diagrams for two linear defects of radii (a) $r_{dU}=r_{dL}=0$ and (b) $r_{dU}=0$ and $r_{dL}=1.2\,\rm{mm}$ including the surrounding rows of radius $r_{dS}=2.5\,\rm{mm}$. We plot the S and the A bands that are used for the control of sound propagation, with the same notation as in \fref{sc_fig2}. The supermodes for the S and A bands are shown for a frequency of $1.9\times 10^5\,\rm{Hz}$. Note that the sound pressure of these supermodes is more confined into the linear defects than in Figs.~\ref{sc_fig2}(a) and (b).}
\label{sc_5defectes}
\end{figure*}

A structure similar to the one described in Section \ref{sc_CMAS} can be designed in order to measure the phase difference between two sound beams. In this case, at the input the two defects are empty of cylinders, being $\theta=\pi/4$ and both the S and the A supermodes spread equally into the two defects, see \fref{sc_fig2}(a). Along the propagation direction, the defect radius of the lower linear defect smoothly increases its size up to $r_{dL}=1.2\,\rm{mm}$, while the upper defect remains without cylinders, $r_{dU}=0$. Thus, at the output of the system, since the radii of the defects are significantly different, $\Delta k\gg\sqrt{\Omega_{UL}\Omega_{LU}}$, $\theta=0$, and the S supermode is localized in the upper waveguide while the A one is localized in the lower defect, as shown in \fref{sc_fig2}(b). Now we consider that two monochromatic sound beam sources of the same frequency are placed at the input of the system, each one in front of one of the two linear defects. If the sources are in phase, only the S supermode will be excited, sound waves will follow adiabatically the supermode along the propagation and at the output of the device there will be only sound intensity in the upper output for which $r_{dU}=0$. On the other hand, if the two sources are in opposite phase, only the A supermode will be excited and sound waves will follow it adiabatically ending up into the lower output for which $r_{dL}=1.2\,\rm{mm}$. In general, one can consider any phase difference $\varphi$ between the input sources. In this case, the normalized initial field for two sources of equal intensity can be expressed as:
\begin{equation}
\psi_{in}=\frac{1}{\sqrt{2}}\left (\begin{array}{c}1\\ e^{i\varphi}\end{array}\right).
\label{sc_inc}
\end{equation}
The S and A supermodes given by the coupled-mode equations, \eref{sc_sim} and \eref{sc_antisim}, at the input where the two defects are equal read:
\begin{equation}
p_{+}(0)=\frac{1}{\sqrt{2}}\left (\begin{array}{c}1\\ 1\end{array}\right),
\label{sc_simin}
\end{equation} 
\begin{equation}
p_{-}(0)=\frac{1}{\sqrt{2}}\left (\begin{array}{c}1\\ -1\end{array}\right).
\label{sc_antisimin}
\end{equation}
Projecting $\psi_{in}$ into the S and A supermodes and taking the modulus square, it is possible to find the sound intensity that excites each supermode, $I_+$ and $I_-$:
\begin{equation}
I_+=|p_{+}(0)^T\ \psi_{in}|^2=\frac{1}{2}(1+\cos{\varphi})
\label{sc_Ip}
\end{equation}
\begin{equation}
I_-=|p_{-}(0)^T\ \psi_{in}|^2=\frac{1}{2}(1-\cos{\varphi}).
\label{sc_Im}
\end{equation}
Since at the output of the system the S supermode corresponds only to sound intensity in the upper linear defect ($I_+=I_{U,out}$), and the A supermode to sound intensity in the lower linear defect ($I_-=I_{L,out}$), it is possible to measure the initial phase difference $\varphi$ between the two sources of the same power by measuring the intensities at the two linear defect outputs:
\begin{equation}
\varphi=\pm\arccos\left(\frac{I_{U,out}-I_{L,out}}{I_{U,out}+I_{L,out}}\right).
\label{sc_def}
\end{equation}
Thus, this structure of linear defects can be used as a phase difference analyzer.

\begin{figure*}
\centerline{
\includegraphics[width=0.75\linewidth]{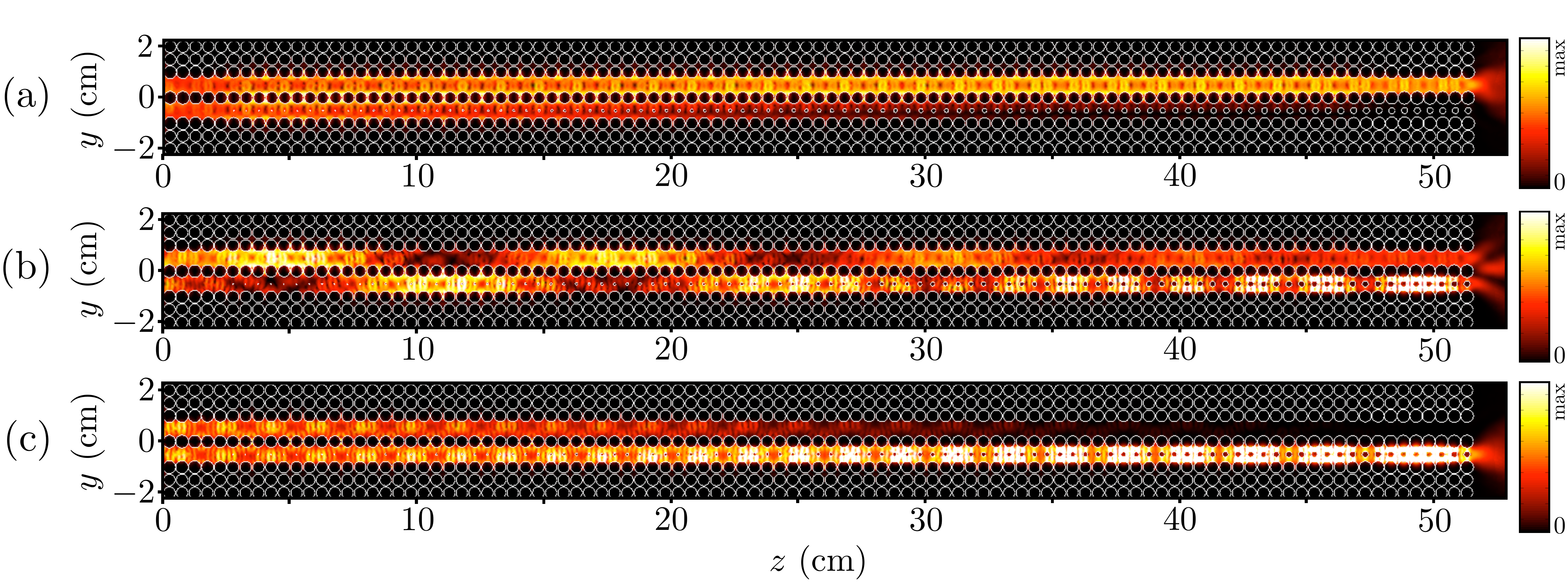}}
\caption{(Color online) Numerical simulations of the intensity of a harmonic sound wave propagating along the SC with two linear defects acting as a phase difference analyzer, with a phase difference between the two input sources of (a) $\varphi=0$, (b) $\varphi=\pi/2$ and (c) $\varphi=\pi$ for a frequency of $1.9\times 10^5\,\rm{Hz}$. The parameter values for the SC are the same as in \fref{sc_supercells}. The upper linear defect consists of a row empty of cylinders whereas the lower linear defect smoothly changes its radius from $r_{dL}=0$ to $r_{dL}=1.2\,\rm{mm}$. The first $10$ columns surrounding the linear defects have radii ranging from $r_{dS}=2.5\,\rm{mm}$ to $r_{dS}=2.25\,\rm{mm}$ and vice-versa for the last $10$ surrounding columns. The two sound sources have a Gaussian transverse pressure profile with a full-width at half maximum of $5\,\rm{mm}$, and they are centered at each linear defect.}
\label{sc_figfase}
\end{figure*}

In order to enhance the working efficiency of the phase difference analyzer, the radius of the cylinders surrounding the linear defects (the cylinders corresponding to the rows above the upper linear defect, between the linear defects and below the lower linear defect) have values smoothly changing from $r_{dS}=2.5\,\rm{mm}$ to $r_{dS}=2.25\,\rm{mm}$ in the first $10$ columns of the SC, and from $r_{dS}=2.25\,\rm{mm}$ to $r_{dS}=2.5\,\rm{mm}$ in the last $10$ columns. In this way, the sound waves are more confined into the waveguides and, therefore, it is easier to correctly excite the S and A supermodes with the two sound sources and, in turn, allows for a better identification of the final intensity at each output. The fact that the supermodes are more confined when they are surrounded by rows of radius $r_{dS}=2.5\,\rm{mm}$ has been checked by comparing the band diagrams and the S and the A supermodes obtained with these extra thicker rows surrounding the linear defects plotted in \fref{sc_5defectes} with the diagrams and supermodes shown in Figs.~\ref{sc_fig2}(a) and (b). These new band diagrams have been determined by considering the surrounding rows as new linear defects of radius $r_{dS}$ and recalculating \eref{sc_coeff2}:
\begin{align}
\rho_{\mathbf{\bar{G}}}^{-1}=&\left(\frac{\rho_h}{\rho_c}-1\right)\frac{2\pi}{9a^2}\notag
\\
&\left[r_{dU}\frac{J_1(|\mathbf{\bar{G}}|r_{dU})}{|\mathbf{\bar{G}}|}e^{i\bar{G}_ya}+r_{dL}\frac{J_1(|\mathbf{\bar{G}}|r_{dL})}{|\mathbf{\bar{G}}|}e^{-i\bar{G}_ya}+\right.\notag
\\
&\left.2r_{dS}\frac{J_1(|\mathbf{\bar{G}}|r_{dS})}{|\mathbf{\bar{G}}|}\left(1/2+\cos(2\bar{G}_ya)\right)+\right.\notag
\\
&\left. 2r_0\frac{J_1(|\mathbf{\bar{G}}|r_0)}{|\mathbf{\bar{G}}|}\sum_{j=3}^{4}{\cos(j\bar{G}_ya)}\right], \,\rm{\ for\ } \mathbf{\bar{G}}\neq0.
\label{sc_coeff2_c}
\end{align}
The expressions for $b_{\mathbf{\bar{G}}}^{-1}$ can be analogously calculated.

\begin{figure}[!ht]
\centerline{
\includegraphics[width=1\linewidth]{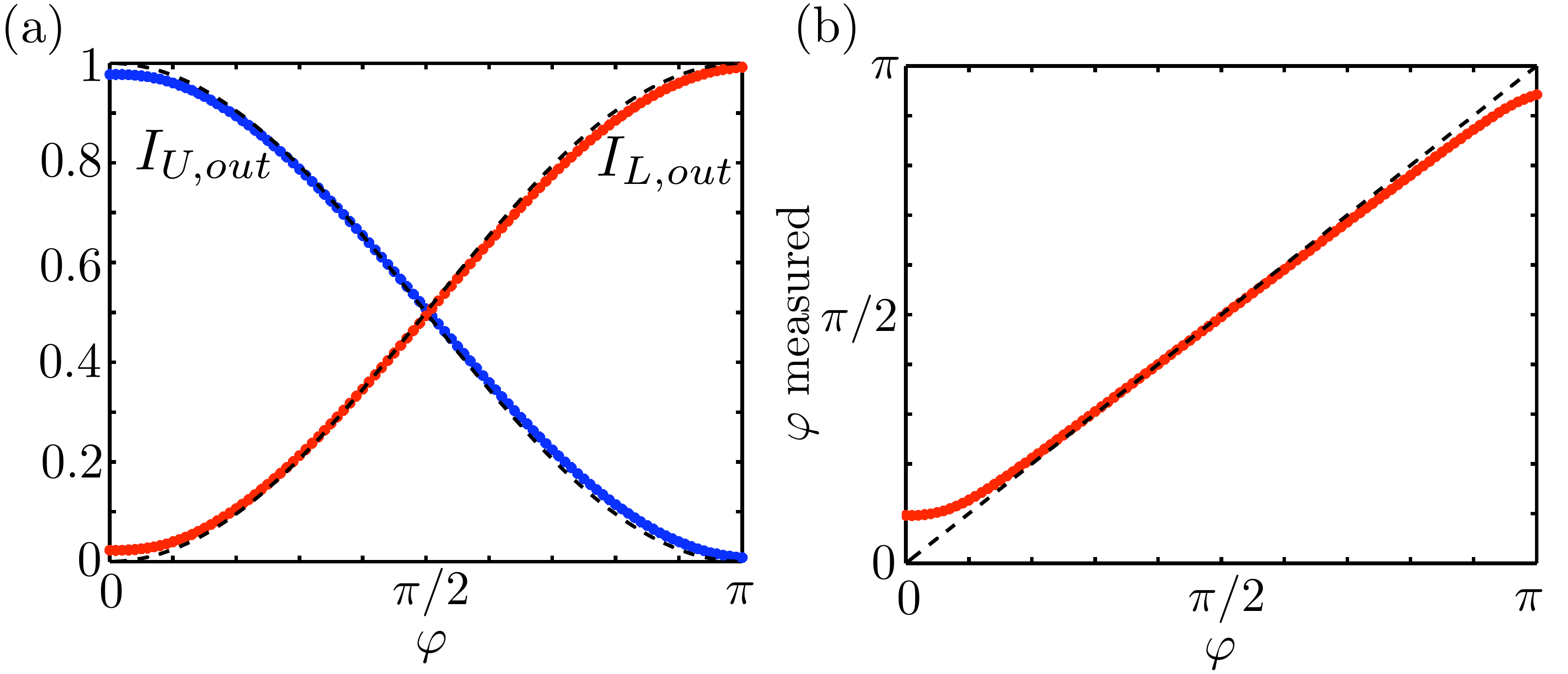}}
\caption{(Color online) (a) Numerically calculated intensity in each waveguide output as a function of the phase difference between the input sources. The results follow those expected from \eref{sc_Ip} and \eref{sc_Im}, represented with a dashed black curve. (b) The solid line represents the measured phase difference from the numerical simulations using \eref{sc_def} as a function of the known phase difference between the sources. The black dashed curve corresponds to the ideal measurement given by \eref{sc_def} together with \eref{sc_Ip} and \eref{sc_Im}.}
\label{sc_figfase2}
\end{figure}

\fref{sc_figfase} shows the results for the sound propagation obtained by numerical integrating \eref{sc_linearized} using the Finite Element method and considering that the sound sources have a Gaussian transverse pressure profile with a full-width at half maximum of $5\,\rm{mm}$, and are centered at each linear defect. We can observe the intensity of the sound waves of frequency $1.9\times 10^5\,\rm{Hz}$ propagating along the designed structure for three different values of the phase difference $\varphi$ between the input sources: (a) $0$, (b) $\pi/2$ and (c) $\pi$. We can clearly see that the intensity at the output depends on the phase difference and that for $\varphi=0$ ($\varphi=\pi$) only the S (A) supermode is excited and followed, while for $\varphi=\pi/2$ both supermodes are equally excited and followed. \fref{sc_figfase2}(a) shows the dependence of the output intensity in each waveguide on the phase difference, which follows the expected behavior (\eref{sc_Ip} and \eref{sc_Im}). Thus, it is possible to obtain the measured phase difference using \eref{sc_def} as a function of the known phase difference introduced between the sources (\fref{sc_figfase2}(b)). The results from the numerical simulations are in good agreement with the analytical ones. Small discrepancies arise only close to $0$ and $\pi$ phase differences. This is due to the impossibility to completely decouple the two linear defects at the input and the output of the system. Nevertheless, a proper calibration of the device would allow for the assignment of a real value of the phase difference to every measured value.

The results shown above correspond to a frequency of $1.9\times 10^5\,\rm{Hz}$, which is approximately in the middle of the frequency range where both S and A supermodes coexist, see Figs.~\ref{sc_fig2}(a) and (b). For this reason, both supermodes are excited with similar intensities. However, for other frequencies surrounding that frequency, one of the supermodes will be excited with slightly more intensity than the other one. Nevertheless, the system can still be used for the measurement of the phase difference between two beams by adjusting the normalization in \eref{sc_simin} and \eref{sc_antisimin}, and also calibrating the obtained outputs.

\section{One-supermode case}\label{sc_onesmode}
Up to now we have studied sound propagation in two parallel linear defects for frequencies in which both S and A supermodes exist. Nevertheless, from Figs.~\ref{sc_fig2}(a) and (b) it is clear that there are frequencies in which only one of the supermodes can propagate. In particular, marked with a gray (red in the online version) shadow in Figs.~\ref{sc_fig2}(a) and (b), there is a quite broad range of frequencies where only the A supermode is present. For these frequencies it is not possible to use the coupled-mode theory since only one supermode is available. However, it is still possible to adiabatically modify and follow the A supermode of the system by smoothly modifying the radius size of the defects along the propagation direction. In fact, the absence of the S supermode significantly relaxes the smoothness condition for the change in radii size of the linear defects along the propagation direction. Therefore, in this range of frequencies it is possible to design much shorter SC structures for applications that only require the following of a single supermode.

\begin{figure}[htb]
\centerline{
\includegraphics[width=\linewidth]{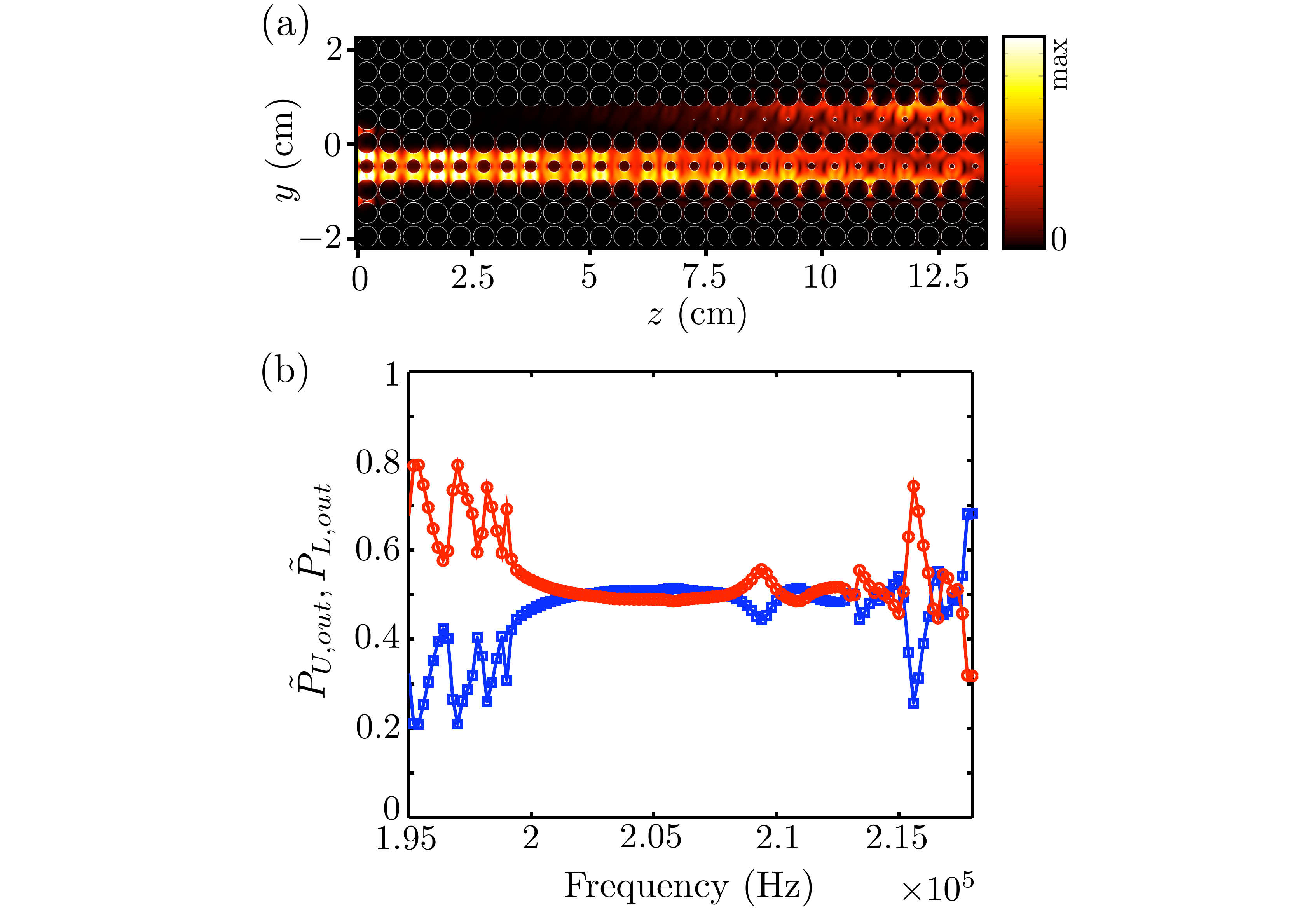}}
\caption{(Color online) (a) Numerical simulation of the intensity of a harmonic sound wave propagating with a frequency of $2.08\times 10^5\,\rm{Hz}$ along the SC with two linear defects for the coherent beam splitter structure. The sound source transverse profile is the same as in \fref{sc_figsplit} but centered at the lower defect. (b) Normalized power at the upper $\tilde{P}_{U,out}$ (red dots) and lower $\tilde{P}_{L,out}$ (blue squares) outputs as a function of the frequency with respect to the total power at the output.}
\label{sc_splitcurt}
\end{figure}

\begin{figure}[!ht]
\centerline{
\includegraphics[width=1\linewidth]{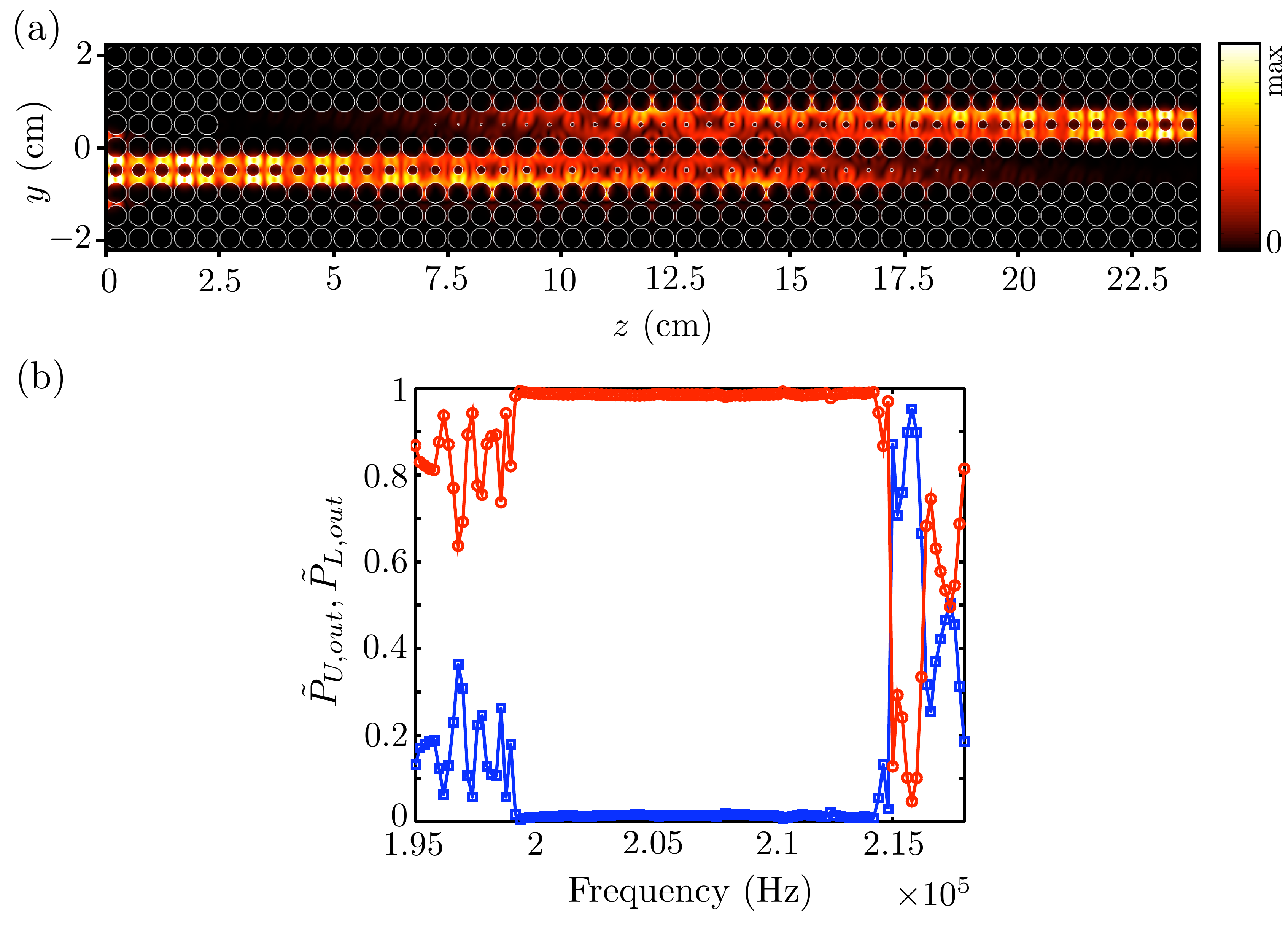}}
\caption{(Color online) (a) Numerical simulation of the intensity of a harmonic sound wave propagating with a frequency of $2.08\times 10^5\,\rm{Hz}$ along the SC with two linear defects for the robust complete transfer structure. The sound source transverse profile is the same as in \fref{sc_figsplit} but centered at the lower defect. (b) Normalized power at the upper $\tilde{P}_{U,out}$ (red dots) and lower $\tilde{P}_{L,out}$ (blue squares) outputs as a function of the frequency with respect to the total power at the output.}
\label{sc_rap_anti}
\end{figure}

This is the case of the coherent multifrequency adiabatic splitter, where by designing a structure with two linear defects starting with very different radii of the defects, $r_{dL}=1.5\,\rm{mm}$ and $r_{dU}=0$, and ending with defects of the same size, $r_{dL}=r_{dU}=0.5\,\rm{mm}$, it is possible to modify the transverse profile of the A supermode from being very localized in one of the defects to equally spread in the two linear defects. In \fref{sc_splitcurt}(a) we show the sound intensity propagation for this case. Additionally, \fref{sc_splitcurt}(b) plots the normalized power with respect to the total power at the upper $\tilde{P}_{U,out}$ and lower $\tilde{P}_{L,out}$ outputs as a function of the frequency. We can see that for frequencies from $2\times 10^5\,\rm{Hz}$ to $2.13\times 10^5\,\rm{Hz}$, which is the range in which only the A supermode is present, there is a $50\%$ splitting of the sound intensity. 
Although the results here show a narrower range of working frequencies and a bit higher oscillations of the total power reaching the output as compared to the one presented in Section \ref{sc_CMAS}, the structure can be significantly shortened (approximately $5$ times with respect to the one presented in Section \ref{sc_CMAS}).

Additionally, given the short length of the splitter, it is possible to design a structure made of two of them, one after the other, with the second one rotated by $180$ degrees with respect to the first one. By doing this, sound waves initially in one of the linear defects are completely transferred to the other linear defect by adiabatically following the A supermode, see \fref{sc_rap_anti}(a). \fref{sc_rap_anti}(b) shows the range of frequencies in which the complete transfer works efficiently, which as expected coincides with the case of the one-supermode splitter. 

\section{Conclusions}\label{sc_CONC}
We have demonstrated for the first time to the best of our knowledge the possibility to apply spatial adiabatic passage processes to the field of sound waves propagation in linear defects. Sonic crystals with two linear defects that change their geometry along the propagation direction are designed in order that the sound waves adiabatically follow a supermode of the system. Two different frequency ranges within the band gap are studied. First, a frequency range in which two supermodes of the system coexist. For this frequency range, two structures have been designed working as a coherent multifrequency adiabatic splitter and as a phase difference analyzer. Second, a range of frequencies in which only one of the supermodes exists. In this range of frequencies, a reduction in length of the designed structures is possible since transitions to any other supermode are strongly suppressed. A coherent multifrequency adiabatic splitter and a coupler have been designed for this range of frequencies. It is important to note that since these applications are based on spatial adiabatic passage processes, they are robust and do not depend on specific parameter values of the physical system. Additionally, the equations for sound propagation in sonic crystals can be directly related to the TE or TM modes for light propagation in two-dimensional systems. Therefore, the applications here discussed could be easily extended to the field of light propagation in photonic crystals with linear defects. Furthermore, the applications described through the coupled-mode equations could also be extended to other kinds of optical waveguides such as standard TIR waveguides.

\section*{Acknowledgements}
The authors gratefully acknowledge financial support through the Spanish MICINN contract FIS2011-23719, the Catalan Government contract SGR2009-00347. R. M.-E. acknowledges financial support from AP2008-01276 (MECD). We also acknowledge L. Maigyte for the given support with the numerical simulations, and V.J.~S\'anchez-Morcillo for his helpful comments.

\end{document}